\documentclass[aps,prb,twocolumn,showpacs,amsmath]{revtex4}
\usepackage{graphicx}
\usepackage{bm}

\begin{document}

\title{Geometrical barriers and the growth of flux domes in thin ideal superconducting disks}

\author{John R.\ Clem}
\affiliation{%
   Ames Laboratory and Department of Physics and Astronomy, \\
   Iowa State University, Ames, Iowa, 50011--3160}

\date{\today}

\begin{abstract} 
When an ideal (no bulk pinning) flat type-II superconducting disk is subjected to a perpendicular magnetic field $H_a$, the first vortex nucleates at the rim when  $H_a = H_0$,  the threshold field, and moves quickly to the center of the disk.  As $H_a$ increases above $H_0$, additional vortices join the others, and together they produce a domelike field distribution of radius $b$.  In this paper I present analytic solutions for the  resulting  magnetic-field and sheet-current-density distributions.  I show how these distributions vary as $b$ increases with $H_a$, and I calculate the corresponding field-increasing magnetization. 
\end{abstract}

%PACS 2006
%74.25.Ha 	Magnetic properties
%74.25.Nf 	Response to electromagnetic fields (nuclear magnetic resonance,
%surface impedance, etc.)
%74.25.Op 	Mixed states, critical fields, and surface sheaths
%74.25.Sv 	Critical currents
%74.78.Bz 	High-Tc films
\maketitle

\section{introduction}

The penetration of magnetic flux into nonellipsoidal superconducting samples is  complex.  The initial entry of magnetic flux in the form of multiply quantized flux tubes in type-I superconductors or singly quantized vortices in type-II superconductors is impeded both by a barrier of geometric origin and by bulk pinning.  The effects of the geometrical barrier are most pronounced in type-I superconductors\cite{Girard71,Huebener72,Clem73,Provost74,Castro99} and in type-II superconductors in which the bulk pinning is unusually low.\cite{Indenbom94,Schuster94,Zeldov94a} Detailed studies of geometrical-barrier effects in strips and long slabs, in which the magnetic-field distribution is two-dimensional and conformal-mapping techniques can be applied, have been carried out for both type-I superconductors\cite{Fortini76,Fortini80} and type-II superconductors\cite{Indenbom94,Schuster94,Zeldov94a,Benkraouda96,Doyle97,Maksimov98,
Benkraouda98,Mawatari01,Brojeny02,Zhelezina02} 

Various authors have carried out numerical calculations of the magnetic-field and current distributions in flat disks of various radii and thicknesses, including the possibility of the formation of a cylindrically symmetric flux dome at the center.\cite{Schweigert99,Deo99,Brandt99,Brandt00,Brandt01,Brojeny03,Berdiyorov05}
When the flux distribution in the disk depends only upon the radial coordinate $\rho$, the problem is two-dimensional, in that the magnetic field can be assumed to depend only upon $\rho$ and the axial coordinate $z$.  However, conformal-mapping techniques can no longer be applied in this geometry.

In this paper I present analytic solutions for the magnetic-field and sheet-current-density distributions for the geometrical-barrier problem in thin superconducting disks.  As in most of the previous theoretical studies, I have ignored the spatial structure on the scale of the spacing between flux tubes or the intervortex spacing and have considered the magnetic flux density averaged over this fine scale, such that the solutions have cylindrical symmetry.  Although I describe magnetic-flux penetration in terms of the entry of vortices into type-II superconducting disks, the theory could be applied with only minor modifications to the entry of flux tubes into type-I superconducting disks.

The paper is organized as follows.  In Sec.\   \ref{small}, I consider the 
initial entry of vortices into a disk and the formation of a flux dome whose radius $b$ is much smaller than the disk radius $a$.
In Sec.\ \ref{large}, I theoretically examine the subsequent entry of magnetic flux as the dome radius increases and $b$ approaches $a$.  
In Sec.\ \ref{magnetization}, I present results for the field-increasing magnetization, and in Sec.\ \ref{summary}, I briefly summarize the main findings of this paper.  Four Appendices contain numerous mathematical details underlying the theory.  Finally, Sec.\ \ref{bardeen} contains some of my recollections of John Bardeen, who is being honored in this issue.

\section{initial growth of small flux domes \label{small}}

Consider an ideal (pinning-free) thin type-II superconducting disk of thickness $d$ and radius $a$  centered on the $z$ axis in the $xy$ plane.  I assume  that either $a \gg d$ if  $d/2 > \lambda,$ where $\lambda$ is the London penetration depth, or $a \gg \Lambda$ if $d/2 < \lambda$, where $\Lambda$ is the Pearl length,\cite{Pearl64} $\Lambda = 2\lambda^2/d.$  In the Meissner (vortex-free) state, when a magnetic field $H_a$ is applied along the $z$ direction,  the magnetic field in  the plane $z=0$ is, as shown in Refs.\ \onlinecite{Brojeny03}, \onlinecite{Mikheenko93}, and \onlinecite{Clem94} and Appendix \ref{Ha},
\begin{eqnarray}
H_{az}(\rho,0)& = &0, \; \rho < a, \\
&=& H_a\Big\{1+\frac{2}{\pi}\Big[\frac{a}{\sqrt{\rho^2-a^2}}-\sin^{-1}\Big(\frac{a}{\rho}\Big)\Big]\Big\}, 
\nonumber \\
&&\;\;\;\;\;\;\rho > a.
\end{eqnarray}
When $\rho \to a$, $H_{az}(\rho,0)$ is dominated by the term with the inverse square-root singularity.

 The corresponding sheet-current density in the disk $(\rho < a)$ is, from Eq.\ (\ref{AKaphi}),
\begin{equation}
K_{a\phi}(x)=-\frac{4H_a}{\pi} \frac{\rho}{\sqrt{a^2-\rho^2}}.
\label{Kaphi}
\end{equation}
Close to the center of the disk,
\begin{equation}
K_{a\phi}(x) \approx -\frac{4H_a \rho}{\pi a}.
\end{equation}

Using the reasoning of Ref.\ \onlinecite{Brojeny03}, I make the simplifying assumption that the first vortex enters when the applied field reaches the value $H_0$, at which the demagnetization-enhanced field at the edge, estimated as 
\begin{equation}
H_{a,edge}\approx H_{az}(a+\delta,0) \approx H_a \frac{2}{\pi}\sqrt{\frac{a}{2\delta}},
\label{Haedge}
\end{equation}
where $\delta$ is the larger of $d/2$ or $\Lambda$, becomes equal to $H_s$, where $H_s = H_{c1}$, the lower critical field, if there is no Bean-Livingston barrier, or $H_s \approx H_c$, the bulk thermodynamic field, if the edge is without defects and thermal activation is negligible. Thus 
\begin{equation}
H_0 \approx H_s \frac{\pi}{2} \sqrt{\frac{2\delta}{a}}.
\label{H0}
\end{equation}
However, to determine precisely when a vortex enters or exits at the edge  is a difficult problem, because these processes are sensitive to details such as the shape and perfection of the edge.\cite{Kupriyanov75,Aslamazov83,
Schuster94,Zeldov94a,Zeldov94b,Benkraouda96,Morozov97,Kuznetsov97,
Doyle97,Kuznetsov98,Deo99,Vodolazov99,Doyle00,Brandt99,Brandt00, Brandt01,Wang02,Elistratov02,Mawatari03, Berdiyorov05}

When $H_a$ is just above $H_0$, vortices enter a pin-free disk and collect in a flux dome at the center, driven by the Lorentz force,
\begin{equation}
F_\rho = K_{a\phi}\phi_0.
\end{equation}
Close to the center of the disk,
\begin{equation}
F_\rho \approx -4H_a \phi_0 \rho/\pi a.
\end{equation}
\begin{figure}%***** Fig.1 ************************
\includegraphics[width=8cm]{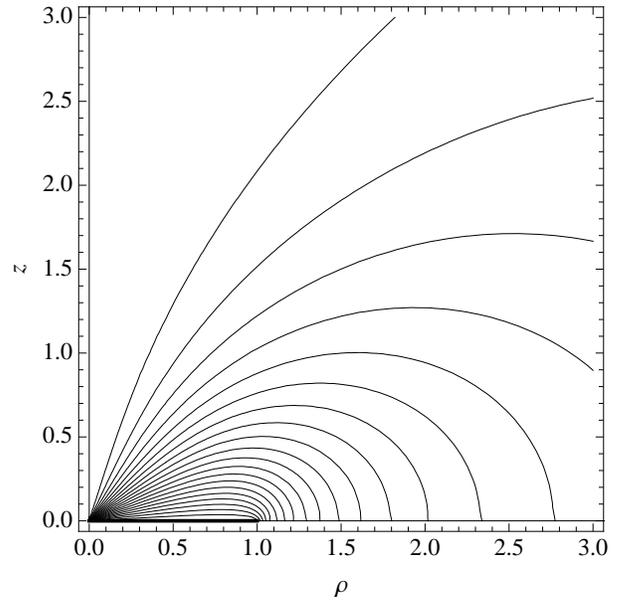}
\caption{%
Contour plot of the stream function $\psi_v(\rho,z)$, where the contours correspond to lines of magnetic field emanating from a flux dome carrying flux $\Phi$ up through the disk at the origin.  The flux $\Phi$ returns around the perimeter of the disk at $\rho = a = 1$. }
\label{psiv}
\end{figure} 

As shown in Fig.\ \ref{psiv}, the vortices in the flux dome generate a magnetic field ${\bm H}_v(\rho,z) = \hat \rho H_{v\rho}(\rho,z) + \hat z H_{vz}(\rho,z),$ which is calculated in Appendix \ref{Hv}.  The  return flux around the rim of the disk partially cancels the field $H_{a,edge}$.  The vortex-generated local field at the outer edge can be estimated as [see Eq.\ (\ref{Hvz>})]
\begin{equation}
H_{v,edge}\approx H_{vz}(a+\delta,0) \approx -\frac{\Phi}{\pi^2 \mu_0 a^2} \sqrt{\frac{a}{2\delta}}.
\label{HvedgeSmall}
\end{equation}

The condition for the next vortex to enter the strip is therefore that 
\begin{equation}
H_{edge} = H_{a,edge}+H_{v,edge}=H_s,
\label{HedgeSmall}
\end{equation}
or 
\begin{equation}
H_a-H_0 = \frac{\Phi}{2 \pi \mu_0 a^2},
\label{diskflux}
\end{equation}
such that the total magnetic flux in the flux dome of a superconducting disk initially grows linearly with the difference $H_a-H_0$.

Next, let us        examine how the radius $b$ of the dome initially varies as a function of $H_a-H_0 $.  The dome-generated magnetic field near the center of the disk looks much like that produced by a distribution of magnetic monopoles.  When $b \ll a$, the dome's magnetic-field distribution can be calculated to good approximation by treating the problem as if the disk were of infinite radius.  As shown in Appendix \ref{Hd}, the magnetic flux $\Phi$ penetrating up through the strip spreads out within $\rho < b$ with the distribution $H_{sz}(\rho,0) = H_d \sqrt{1-\rho^2/b^2}$.  The corresponding dome-generated magnetic field tangent to the top surface within $\rho < b$ for a small dome is 
\begin{equation}
H_{s\rho}(\rho,0+) =  \frac{\pi H_d \rho}{4 b}.
\end{equation}
The field on the bottom surface has the same behavior but of opposite sign, such that the dome-generated sheet-current density behaves  as
\begin{equation}
K_{s\phi}(\rho) = \frac{\pi H_d \rho}{2 b}.
\end{equation} 
However, the net sheet-current density $K_\phi(x)$ inside the dome must vanish in order that there be zero net Lorentz force on any vortex within the dome.  Setting $K_\phi(\rho) = K_{a\phi}(\rho) + K_{s\phi}(\rho) = 0$ yields for a small dome
\begin{equation}
H_d = \frac{8}{\pi^2} H_a \frac{b}{a}.
\label{Hddisk}
\end{equation}
Outside the dome, assuming $b \le \rho \ll a$, the results of Appendix \ref{Hd} can be used to show that 
\begin{equation}
K_\phi(\rho) = -H_d\Big[\frac{\rho}{b}\Big(\frac{\pi}{2}-\sin^{-1}\frac{b}{\rho}\Big) 
+\frac{\sqrt{\rho^2-b^2}}{\rho}\Big].
\end{equation}
Since $K_\phi(\rho) < 0$, the radial Lorentz force on any vortex outside the dome obeys $F_\rho(\rho) =K_\phi(\rho) \phi_0 < 0$, such that all vortices are confined within the dome. 

\begin{figure}%***** Fig.2 ************************
\includegraphics[width=8cm]{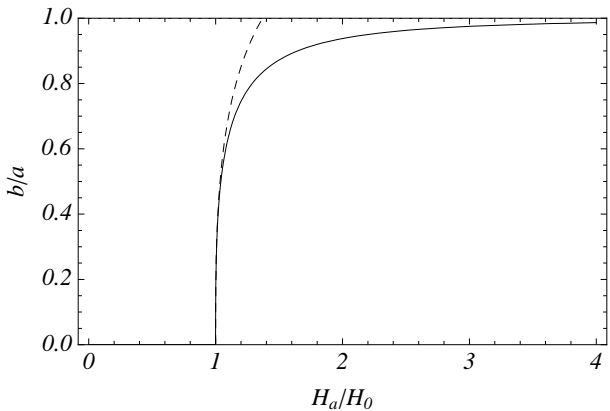}
\caption{%
Dome radius vs increasing applied field, expressed as $b/a$ vs $H_a/H_0$.  The dashed curve shows the small-dome approximation of Eq.\ (\ref{boveraSmall}), and the solid curve shows the theoretical results for  domes of arbitrary radius, obtained from Eq.\ (\ref{boveraLarge}).}
\label{betavsha}
\end{figure}

The magnetic flux contained within a small dome is 
\begin{equation}
\Phi = 2\pi\mu_0 \int_0^b H_{sz}(\rho) \rho d\rho = 2\pi \mu_0 H_d b^2/3.
\label{Phidisk}
\end{equation}
Combining this result with Eqs.\ (\ref{diskflux}) and (\ref{Hddisk}) leads to the condition
\begin{equation}
\frac{H_a-H_0}{H_a} = \Big(\frac{8}{3 \pi^2}\Big) \Big(\frac{b}{a}\Big)^3
=0.270 \Big(\frac{b}{a}\Big)^3,
\label{bdisk}
\end{equation}
or
\begin{equation}
\frac{b}{a} = \Big(\frac{3\pi^2}{8}\Big)^{1/3} \Big(\frac{H_a-H_0}{H_a}\Big)^{1/3}
= 1.547 \Big(\frac{H_a-H_0}{H_a}\Big)^{1/3},
\label{boveraSmall}
\end{equation} 
when $b/a \ll 1$.
Numerical calculations of the dome radius carried out as in Ref.\ \onlinecite{Brojeny03} are in agreement with these results. 
The dashed curve in Fig.\ \ref{betavsha} shows a plot of $b/a$ given by the small-dome approximation of Eq.\ (\ref{boveraSmall}) for an increasing applied field $H_a$.

If the applied field $H_a$ is reduced after a small dome has appeared at the center, the magnetic flux $\Phi$ in the dome remains constant because the edge conditions no longer permit the entry of vortices, and since $K_\phi(\rho) < 0$ for $b < \rho < a$, any vortex that strays out of the dome is forced back in.  Thus as $H_a$ decreases, the dome radius $b$ expands and the field $H_d$ at the center decreases.  According to Eqs.\ (\ref{Hddisk}) and (\ref{Phidisk}),
\begin{equation}
b = \Big(\frac{3 \pi \Phi a}{16 \mu_0 H_a}\Big)^{1/3}
\end{equation}
and 
\begin{equation}
H_d = \Big(\frac{96 \Phi H_a^2}{\pi^5 \mu_0 a^3}\Big)^{1/3}.
\end{equation}

\section{Arbitrary flux dome radius \label{large}}

As the applied field $H_a$ increases and additional vortices move to the center, the radius $b$ of the dome expands, and the small-dome approximations used in the previous section are no longer valid, because they assume that $b \ll a$.  In particular, the small-$b$ dome shape, $H_{sz}(\rho,0) = H_d \sqrt{1-\rho^2/b^2}$, is no longer valid.  Appendix  \ref{Hc}  contains the details of the magnetic-field and sheet-current distributions resulting from an array of vortices carrying magnetic flux 
$\Phi$ up through an annulus of radius $\rho_c < a$.  These results can be used to obtain the Green's functions that generate the magnetic-field and sheet-current-density distributions produced in response to a large flux dome characterized by an arbitrary field distribution 
$H_{dz}(\rho',0)$ confined to the region $\rho' < b < a$.  We need only to replace $\rho_c$ by $\rho'$ and $\Phi$ by 
$2\pi \mu_0 H_{dz}(\rho',0)\rho'd\rho'$ and then integrate over $\rho'$ from 0 to $b$.  The resulting expression for the $z$ component of the dome-generated magnetic field for $\rho > a$ is
\begin{equation}
H_{dz}(\rho,0)=-\frac{2}{\pi}\int_0^b
\frac{H_{dz}(\rho',0)\rho'\sqrt{a^2-\rho'^2}}{(\rho^2-\rho'^2)\sqrt{\rho^2-a^2}}d\rho'.
\label{Hdzlarge}
\end{equation}
This expression has the property that 
\begin{equation}
2\pi\int_a^\infty H_{dz}(\rho,0)\rho d\rho =-2\pi\int_0^b H_{dz}(\rho,0)\rho d\rho,
\end{equation}
which is equivalent to the statement that the return flux through the space $\rho > a$ is the negative of the magnetic flux up through the dome $\rho < b$.  Similarly, the resulting expression for the dome-generated sheet-current density is
\begin{equation}
K_{d\phi}(\rho)=2\int_0^bH_{dz}(\rho',0){\cal G}(a,\rho',\rho)\rho'd\rho',
\end{equation}
where ${\cal G}(a,\rho',\rho)$ is defined in Appendix \ref{Hc}.

Because the disk is in a perpendicular applied field $H_a$, which induces its own sheet-current density $K_{a\phi}(\rho)$ given in Eq.\ (\ref{Kaphi}), the net sheet-current density is $K_\phi(\rho) = K_{a\phi}(\rho)+K_{d\phi}(\rho)$.  For the vortices in the dome to be in equilibrium, the net Lorentz force on any vortex within the dome must obey $F_\rho(\rho) = K_\phi(\rho) \phi_0=0$.  The field distribution $H_{dz}(\rho,0)$ inside the dome therefore must obey the integral equation
\begin{equation}
\int_0^bH_{dz}(\rho',0){\cal G}(a,\rho',\rho)\rho'd\rho'=\frac{2H_a\rho}{\pi\sqrt{a^2-\rho^2}}
\label{Hintequation1}
\end{equation}
for all $\rho<b$.

Another equivalent integral equation can be obtained by noting that the condition $K_\phi(\rho)=2H_\rho(\rho,0+) = 0$ for $\rho < b$ is equivalent to the condition that $\partial \phi(\rho,0+)/\partial \rho = 0$ or
\begin{eqnarray}
\phi(\rho,0+)&=&\phi_d(\rho,0+)+\phi_a(\rho,0+)\nonumber \\
&=&\phi_d(0,0+)+\phi_a(0,0+),
\label{constantphi}
\end{eqnarray}
where $\phi_a(\rho,0+)=\phi_{a2}(\rho,0+)$ is given in Eq.\ (\ref{phia2}), and $\phi_d(\rho,0+)$ can be obtained from $\phi_c(\rho,0+)$ in the same way that $K_{d\phi}(\rho)$ was obtained from $K_{c\phi}(\rho)$:
\begin{equation}
\phi_d(\rho,0+)=\int_0^bH_{dz}(\rho',0){\cal F}(a,\rho',\rho)\rho'd\rho'.
\label{phid}
\end{equation}
With the help of Eq.\ (\ref{curlyF0}), Eq.\ (\ref{constantphi}) can be expressed as the following integral equation, from which $H_{dz}(\rho,0)$ can be determined: 
\begin{eqnarray}
\int_0^bH_{dz}(\rho',0)\Big[\frac{2}{\pi}\cos^{-1}\frac{\rho'}{a}
-{\cal F}(a,\rho',\rho)\rho'\Big]d\rho' \nonumber \\
=\frac{2H_a}{\pi}(a-\sqrt{a^2-\rho^2}), \; \rho < b.
\label{Hintequation2}
\end{eqnarray}
Equation (\ref{Hintequation1}) can be recovered by differentiating Eq.\ (\ref{Hintequation2}) with respect to $\rho$.  These equations are solved exactly by $H_{dz}(\rho,0)=(8/\pi^2)(b/a)H_a\sqrt{1-\rho^2/b^2}$ in the limit as $b/a \to 0$ and  by $H_{dz}(\rho,0)=H_a$  when $b = a$.  However, I have been unable  to find analytic solutions  for $H_{dz}(\rho,0)$ for arbitrary values of $b/a$, unlike the  situation for a flux dome in a superconducting strip, for  which the integral equation can be inverted exactly for all dome sizes.\cite{Benkraouda96} 

On the other hand, numerical solutions  of Eq.\   (\ref{Hintequation2}) reveal that  the functional form of  $H_{dz}(\rho,0)$ is within a few percent of 
\begin{equation}
H_{dz}(\rho,0)=H_d \sqrt{\frac{1-\rho^2/b^2}{1-\rho^2/a^2}},
\label{Hdzapprox}
\end{equation}  
similar to the behavior of the dome  shape in a  strip,\cite{Zeldov94a,Benkraouda96} except that here  the value of $H_d$ increases from $(8/\pi^2)(b/a)H_a$  for  $b/a\ll 1$ to $H_a$ as  $b/a \to 1$.  Thus, the value of $H_d$ can be obtained to good accuracy by substituting Eq.\ (\ref{Hdzapprox}) into Eq.\ (\ref{Hintequation2}) and evaluating it numerically at $\rho = b$.  The results for $H_d$ vs $\beta = b/a$ for $0 \le \beta \le 1$ can be fit to 0.1\% by 
\begin{equation}
H_d = H_a f_1(\beta),
\label{Hdfit}
\end{equation}
where
\begin{equation}
f_1(\beta) = \beta (\frac{8}{\pi^2} + 0.165 \beta^2 - 0.028 \beta^4 + 0.052 \beta^6).
\label{f1}
\end{equation}

Recall that the growth of flux domes when the dome radius $b$  obeys $\beta=b/a \ll 1$ is determined by Eqs.\ (\ref{HvedgeSmall}) and (\ref{HedgeSmall}).  For larger domes, the dome-generated local field at the outer edge can be obtained with good accuracy from Eqs.\ (\ref{Hdzlarge}) and (\ref{Hdzapprox}) as 
\begin{eqnarray}
H_{d,edge}\!\!&\approx &H_{dz}(a+\delta,0) \nonumber \\
&\approx &
-\frac{2H_d}{\pi b}\sqrt{\frac{a}{2\delta}}\int_0^b
\frac{\rho'\sqrt{b^2-\rho'^2}}{(a^2-\rho'^2)}d\rho' \nonumber \\
&=&-\frac{2H_d}{\pi}f_2(\beta)\sqrt{\frac{a}{2\delta}},
\label{HdedgeLarge}
\end{eqnarray}
where
\begin{equation}
f_2(\beta)=1-
\frac{\sqrt{1-\beta^2}}{\beta}\tan^{-1}\frac{\beta}{\sqrt{1-\beta^2}}.
\end{equation}

The condition for the next vortex to enter the strip is therefore that 
\begin{equation}
H_{edge} = H_{a,edge}+H_{d,edge}=H_s.
\label{HedgeLarge}
\end{equation}
Combining Eqs.\ (\ref{Haedge}), (\ref{H0}), (\ref{Hdfit}), (\ref{HdedgeLarge}), and (\ref{HedgeLarge}), we obtain for $H_a > H_0$ 
\begin{equation}
\frac{H_a}{H_0}=\frac{1}{1-g(\beta)},
\label{boveraLarge}
\end{equation}
where $\beta = b/a$ and 
\begin{equation}
g(\beta)= f_1(\beta)f_2(\beta).
\end{equation}
A plot of the reduced dome radius $\beta=b/a$ vs 
$H_a/H_0$, obtained from Eq.\ (\ref{boveraLarge}), is plotted as the solid curve in Fig.\ \ref{betavsha}.

From Eqs.\ (\ref{Hdzapprox}) and (\ref{Hdfit}) we obtain the corresponding approximation to the total magnetic flux up through the disk into the dome,
\begin{equation}
\Phi\! = \!2\pi\!\!\int_0^b\!\!\!\!H_{dz}(\rho,0)\rho d\rho \!=\! H_d f_3(\beta)\!=\!H_a f_1(\beta)f_3(\beta),
\label{PhiLarge}
\end{equation}
where $\beta = b/a$ and
\begin{equation}
f_3(\beta)=\Big[1-(1-\beta^2)\frac{\tanh^{-1}\beta}{\beta}\Big].
\label{f3}
\end{equation}

Consider the situation when an increasing applied field reaches the value $H_{a1}$, producing a dome of magnetic flux $\Phi_1$, radius $b_1$, and field $H_{d1}$ at the center, where these quantities are related via Eqs.\ (\ref{Hdzapprox}), (\ref{Hdfit}), (\ref{PhiLarge}), and (\ref{f3}).  Suppose that the applied field $H_a$ is now reduced below $H_{a1}$.  The magnetic flux $\Phi$ in the dome remains constant at the value $\Phi_1$ because the edge conditions no longer permit the entry of vortices, and since $K_\phi(\rho) < 0$ for $b < \rho < a$, any vortex that strays out of the dome is forced back in.  Because both $f_1(\beta)$ and $f_3(\beta)$ are increasing functions of $\beta$, as $H_a$ decreases, the dome radius $b$ expands and the field $H_d$ at the center decreases. The new values of $\beta$ and $H_d$ can be obtained from Eq.\ (\ref{PhiLarge}).

\section{Field-increasing magnetization \label{magnetization}}

The above results can be used to calculate the magnetic moment per unit volume, $M_z = m_z/\pi a^2 d$, which I shall refer to as the magnetization, where the magnetic moment is\cite{Jackson62}
\begin{equation}
m_z = \pi \int_0^a \rho^2 K_\phi(\rho) d\rho.
\end{equation}
Because  $K_\phi(\rho) = 2H_\rho(\rho,0+) = -2\partial \phi(\rho,0+)/\partial \rho$ and $\phi(a,0+)=0$, partial integration yields 
\begin{equation}
m_z = 4\pi \int_0^a \rho \phi(\rho,0+) d\rho,
\end{equation}
where $\phi(\rho,0+)=\phi_a(\rho,0+)+\phi_d(\rho,0+)$ and the corresponding contributions to $M_z$ are $M_{az}$ and $M_{dz}$.  

Since $\phi_a(\rho,0+)=\phi_{a2}(\rho,0+)$, given in Eq.\ (\ref{phia2}), we easily obtain the negative diamagnetic contribution arising from the Meissner response to the applied field,\cite{Clem94}
\begin{equation}
M_{az}=-\chi_0 H_a,\;\chi_0 = 8a/3\pi d.
\label{Maz}
\end{equation}

A positive contribution $M_{dz}$ due to vortices in the flux dome is obtained with the help of Eqs.\ (\ref{phid}) and (\ref{Hdzapprox}):
\begin{equation}
M_{dz}=3\chi_0 H_d \int_0^{\theta_b}\sin\theta\sqrt{1-\frac{\sin^2\theta}{\sin^2\theta_b}}{\cal H}(\theta)d\theta,
\label{Mdz}
\end{equation}
where $\theta_b = \sin^{-1}(b/a)$,
\begin{eqnarray}
{\cal H}(\theta) &=& \!{\bm E}(\sin\theta) \nonumber \\
-&&\!\!\!\!\!\!\!\!\!\!\!\!\int_0^{\pi/2}\frac{2\cos\theta[{\bm E}(\cos\phi)\!-\!\sin^2\phi{\bm K}(\cos\phi)]}{\pi\cos\phi(1-\sin^2\theta\cos^2\phi)}d\phi,
\end{eqnarray}
and ${\bm K}(\kappa)$ and ${\bm E}(\kappa)$ are complete elliptic integrals of the first and second kind of modulus $\kappa$.  
Since ${\cal H}(0) = 1,$ when $\beta = b/a \ll 1$, $M_{dz} \approx \chi_0 H_d \beta^2 =  \chi_0 H_a (8/\pi^2) \beta^3 = 3\chi_0(H_a-H_0)$. 
In the limit as $b/a\to 1$, $\theta_b  \to \pi/2$, $H_d \to H_a$, the integral in Eq.   (\ref{Mdz}) yields 1/3, and $M_{dz} \to \chi_0 H_a$.

\begin{figure}%***** Fig.3 ************************
\includegraphics[width=8cm]{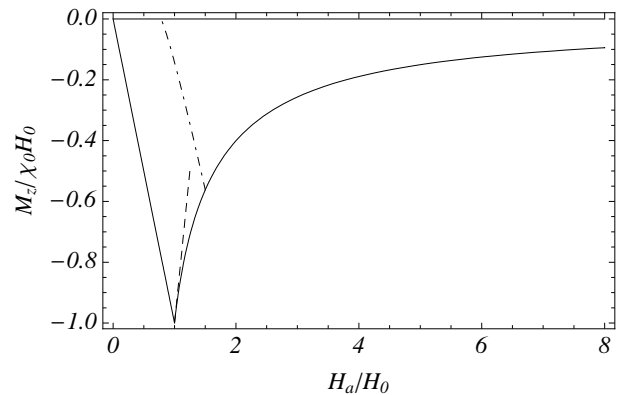}
\caption{%
Magnetic moment per unit volume $M_z$ in units of $\chi_0H_0$ vs increasing applied field $H_a$ in units of $H_0$ calculated from Eqs.\ (\ref{Maz}) and (\ref{Mdz}) (solid).  
The short dashed line shows the behavior $M_z = -\chi_0(3H_0-2H_a)$, which holds just above $H_0$ when $b/a \ll 1$.
The dot-dashed curve shows the reversible magnetization achieved after the applied magnetic field is first increased to 1.5$H_0$ and then decreased; the magnetic flux in the dome remains constant as $H_a$ decreases and the dome radius $b$ increases, thereby decreasing the area that can carry a screening current.}
\label{Mzplot}
\end{figure}

The total magnetic moment per unit volume $M_z=M_{az}+M_{dz}$, numerically  calculated as above with the help of Eqs.\ (\ref{Hdfit}),  (\ref{f1}), and (\ref{boveraLarge}), is shown as the solid curve in Fig.\ \ref{Mzplot}.  Just above $H_0$, $M_z \approx \chi_0(2H_a-3H_0)$, as shown by the dashed line. In the limit as $H_a/H_0\to\infty$, $M_{dz} \to \chi_0 H_a$, nearly canceling $M_{az}$, such that $M_z \to 0$.  However, all the above analysis is valid only when $H_a < H_{c2}$, so that the superconductor remains in the mixed state.

Suppose the increasing-field magnetization follows the solid curve in Fig.\ \ref{Mzplot} in an increasing field $H_a$ up to the value $H_{a1}$, producing a dome containing magnetic flux $\Phi_1$.  As described at the end of Sec. \ref{large}, if the applied field $H_a$ is then reduced, the magnetic flux $\Phi$ remains constant, but $\beta = b/a$ increases and $H_d$ decreases, and Eq.\ (\ref{PhiLarge}) can be used to obtain both $\beta$ and $H_d$ as functions of $H_a$.  In turn, these values can be substituted into 
Eq.\ (\ref{Mdz}) to obtain $M_z = M_{az}+M_{dz}$ as $H_a$ decreases below $H_{a1}$.  This procedure has been used to calculate the resulting  dot-dashed portion of the magnetization curve shown in Fig.\ \ref{Mzplot}, where $H_{a1} = 1.5H_0$ has been used as an example.  This portion of the magnetization curve is reversible, provided that $H_a$ changes slowly and $H_a$ remains above the critical exit field $H_{ex}$, where $b$ is sufficiently close to $a$ that vortices escape from the sample.

\section{Summary \label{summary}}

In this paper I have presented analytic calculations describing the evolution of the flux dome of radius $b$ produced at the center of a thin ideal type-II superconducting disk of radius $a$ when an increasing perpendicular magnetic field $H_a$ is applied.  This phenomenon, not observable in samples with strong bulk pinning, is due to the geometrical barrier.  The theory is exact in the limits $b/a \ll 1$ and $b/a \to 1$.  For intermediate values of $b/a$, this theory is approximate, but the results should be accurate to within  one or two percent.

I have found that $b/a$ varies as $(H_a-H_0)^{1/3}$ when $H_a$ is just above the threshold field $H_0$, but I also have shown how to calculate $b/a$ with good accuracy for all values of $H_a$ above $H_0$.  In addition, I have shown how to obtain the magnetic field $H_d$ at the center of the dome corresponding to each value of $\beta = b/a$. Both $b$ and $H_d$ should be accessible experimentally using scanning techniques with high field sensitivity and spatial resolution. 

I also have presented calculations of the magnetization (or more precisely, the magnetic moment per unit volume) of an ideal type-II superconducting disk as a function of $H_a$ when the magnetization is dominated by the geometrical barrier.  This quantity is also readily observable experimentally.

\section{John Bardeen \label{bardeen}}

Because this issue celebrates the 100th anniversary of John Bardeen's birth, Vladimir Kresin has asked me to add some personal recollections.  When I was in my final year as a graduate student at the University of Illinois in Urbana, Leo Kadanoff, my first advisor, went off on sabbatical leave and left me in the care of John Bardeen.  I periodically reported my progress to Professor Bardeen and remember well that wonderful day when he said these precious words, ``I think you have enough here for a thesis."  I diligently went to work writing my Ph.D. thesis on anisotropy of the superconducting energy gap, and  as I completed the chapters, I sent them to various scientific institutions around the world where John Bardeen was visiting, so that he could peruse what I had written.   
(By the way, my usually tolerant wife Judy reached the end of her patience, phoned me at my office, and instructed me to  come  home when I was working even on the 4th of July.)  

Although John had a quiet, modest manner, he did enjoy entertaining at his home at parties, which I expect were mostly organized by his wife Jane, an outgoing, gracious hostess.  At such a party I recall a funny moment, which John also greatly enjoyed, when a fellow grad student Richard Craig ran around the back yard trying to catch a rabbit.  

John Bardeen was a soft-spoken man, and as students taking classes from him, we knew we had to sit in one of the front rows so that we could hear the pearls of wisdom he could pass on to us.  I also remember when we grad students would sit through a long, obscure theory seminar given by a visitor, only to hear John Bardeen, during  the question and answer period  at the end, summarize in one clear sentence the speaker's main  conclusions and why they were important. 

In addition to his brilliance  as a  scientist, John Bardeen had an unusual trait that it took me some time to get used to.
In many conversations  I have  with others, our sentences are overlapping, with one person starting  to  speak before the  other has  finished.   I suppose this is  because most  of us process input and output  simultaneously.   (Of course, the disadvantage  of this conversational mode is that we too  often speak  before thinking.)  I eventually learned that it was  not this  way  with John Bardeen.   When I said  something to him  or asked him a question, he took some time  to  think about his reply.  While he was  thinking, his  input channel was  turned off, and if I spoke during this  interval, he could not hear me.  I eventually learned to wait patiently for his reply, after which our conversation would continue.
   
It was a privilege  to have known  John Bardeen, whose scientific contributions have had an enormous impact on both science and  technology, and I appreciate this opportunity  to help celebrate his 100th   birthday.

\begin{acknowledgments}
I thank S. J. Bending for posing questions that
stimulated my work on this problem, and I thank A. A. Babaei Brojeny and V. J. Kogan for helpful advice.  Work at the Ames Laboratory was
supported by the Department of Energy - Basic Energy Sciences under Contract
No. DE-AC02-07CH11358.
\end{acknowledgments}

\appendix 

\section{Meissner response of a disk in a perpendicular field \label{Ha}}

This Appendix contains a detailed description of the magnetic field in the vicinity of a superconducting disk of radius $a$ centered at the origin when it is subjected to a perpendicular applied field $H_a$.  In the absence of the disk, a uniform magnetic field in the $z$ direction produced by distant sources can be represented by a potential $\phi_{a1}(z) = -H_a z$ and a stream function $\psi_{a1}(\rho) = -H_a\rho^2/2$.  The magnetic field ${\bm H}_{a1} = \hat \rho H_{a1\rho} + \hat z H_{a1z}$ can be expressed as $H_{a1z} = -\partial \phi_{a1}/\partial z = -(1/\rho)\partial \psi_{a1}/\partial \rho = H_a$, and $H_{a1\rho} = -\partial \phi_{a1}/\partial \rho = (1/\rho)\partial \psi_{a1}(\rho)/\partial z = 0.$ 

The superconducting disk responds with induced screening currents, which generate a magnetic field that cancels the $z$ component of the applied field inside the disk.  The  magnetic field resulting from the induced screening currents,  ${\bm H}_{a2} = \hat \rho H_{a2\rho} + \hat z H_{a2z}$, is described by the potential $\phi_{a2}(\rho,z)$ and stream function $\psi_{a2}(\rho,z)$ via  $H_{a2z} = -\partial \phi_{a2}/\partial z = -(1/\rho)\partial \psi_{a2}/\partial \rho$, and $H_{a2\rho} = -\partial \phi_{a2}/\partial \rho = (1/\rho)\partial \psi_{a2}(\rho)/\partial z,$  where $\nabla^2 \phi_{a2} = 0$.
Here and in the rest of the Appendix, I use a formulation and method described by Lamb.\cite{Lamb45}    The solutions for $z>0$ are 
\begin{eqnarray}
\phi_{a2}(\rho,z)&=&\int_0^\infty f_a(k) J_0(k\rho) e^{- kz} dk,\\
\psi_{a2}(\rho,z)&=&-\int_0^\infty f_a(k) J_1(k\rho) \rho e^{- kz} dk,\\
H_{a2z}(\rho,z)&=&\int_0^\infty f_a(k)J_0(k\rho) e^{-kz}kdk,
\label{Ha2rhoz} \\
H_{a2\rho}(\rho,z)&=&\int_0^\infty f_a(k)J_1(k\rho) e^{-kz}kdk, 
\label{Ha2rhorhoz}\\
f_a(k) &=& \frac{2H_a}{\pi} \frac{d}{dk}\Big(\frac{\sin ka}{k}\Big),
\end{eqnarray}
In the plane $z=0$, 
\begin{eqnarray}
\phi_{a2}(\rho,0+)&=& -\frac{2 H_a}{\pi} \sqrt{a^2-\rho^2}, \; \rho \le a, 
\label{phia2} \\
&=& 0, \;\rho \ge a,\\
\psi_{a2}(\rho,0+)&=& \Big(\frac{H_a \rho^2}{\pi}\Big) \frac{\pi}{2}, \; \rho \le a,
\label{psia2rho0<}  \\
&=& \Big(\frac{H_a \rho^2}{\pi}\Big)\Big[\sin^{-1}\frac{a}{\rho}
-\frac{a\sqrt{\rho^2-a^2}}{\rho^2}\Big], 
\nonumber \\ 
&&\;\;\;\;\;\;\;\;\;\;\;\;\;\;\;\;\;\;\;\;\rho \ge a,
\label{psia2rho0>} \\
H_{a2z}(\rho,0)&= &-H_a, \; \rho < a,\\
&=& \frac{2 H_a}{\pi}\Big(\frac{a}{\sqrt{\rho^2-a^2}}-\sin^{-1}\frac{a}{\rho}\Big), \nonumber \\
&&\;\;\;\;\;\;\;\;\;\;\;\rho > a,\\ 
H_{a2\rho}(\rho,0+)&= &-\frac{2 H_a \rho}{\pi\sqrt{a^2-\rho^2}}, 
\label{Ha2rho<} \; \rho < a,\\
&=& 0,\;\rho > a.
\label{Ha2rho>}
\end{eqnarray}

The overall field distribution, including both the applied field and the Meissner response of the disk, is ${\bm H}_{a}(\rho,z) = {\bm H}_{a1} + {\bm H}_{a2}(\rho,z)=\hat \rho H_{a\rho}(\rho,z) + \hat z H_{az}(\rho,z)$, which is derivable from the sum of the two potentials, $\phi_a(\rho,z) = \phi_{a1}(z)+\phi_{a2}(\rho,z)$, or the sum of the two stream functions, $\psi_a(\rho,z) = \psi_{a1}(\rho)+\psi_{a2}(\rho,z)$, via $H_{az}= -\partial \phi_a(\rho,z)/\partial z = -(1/\rho)\partial \psi_a(\rho,z)/\partial \rho$ and $H_{a\rho}= -\partial \phi_a(\rho,z)/\partial \rho=(1/\rho)\partial \psi_a(\rho,z)/\partial z$. 
In the plane of the disk,\cite{Mikheenko93}
\begin{eqnarray}
H_{az}(\rho,0) &= &0, \; \rho < a,\\
&=& H_a+ \frac{2 H_a}{\pi}\Big(\frac{a}{\sqrt{\rho^2-a^2}}-\sin^{-1}\frac{a}{\rho}\Big), \nonumber \\&&\;\;\;\;\;\rho > a, 
\end{eqnarray}
and $H_{a\rho}(\rho,0+) = H_{a2\rho}(\rho,0+)$, which is given in Eqs.\ (\ref{Ha2rho<}) and (\ref{Ha2rho>}).  Since $H_{a\rho}(\rho,0-)=-H_{a\rho}(\rho,0+)$, the induced sheet-current density in the disk is
 ($\rho < a$) is 
\begin{equation}
K_{a\phi}(\rho)=2H_{a\rho}(\rho,0+)=-\frac{4 H_a \rho}{\pi\sqrt{a^2-\rho^2}}.
\label{AKaphi}
\end{equation}
It is easily shown that 
\begin{equation}
\int_0^a [K_{a\phi}(\rho)/2\rho]d\rho = H_{a2z}(0,0) = -H_a,
\end{equation}
as required by the Biot-Savart law.

\section{Magnetic flux $\Phi$ at the disk's center \label{Hv}}

Consider next the magnetic field and induced current density when a vortex carries magnetic flux $\Phi=\phi_0 = h/2e$ up through the center of the disk, or when the diameter $b$ of a flux dome carrying magnetic flux $\Phi = N \phi_0$ up through the disk is much smaller than the radius $a$ of the disk.   Neglecting  magnetic structure in the dome on the length scale of $b$, $\lambda$, or $\Lambda$, the $z$ component of the magnetic field generated by the vortices is
\begin{eqnarray}
H_{vz}(\rho,0) &= &\frac{\Phi}{2\pi\mu_0\rho}\delta(\rho),\; \rho < a, \label{Hvz<} \\
&=&-\frac{\Phi a}{\pi^2\mu_0\rho^2 \sqrt{\rho^2-a^2}}, 
\label{Hvz>}\; \rho >a.
\end{eqnarray}
Both $H_{vz}(\rho,z)$ and the corresponding radial component $H_{v\rho}(\rho,z)$ can be expressed in terms of the  magnetic potential $\phi_{v}(\rho,z)$ and stream function $\psi_v(\rho,z)$ via $H_{vz}(\rho,z)= -\partial \phi_v(\rho,z)/\partial z = -(1/\rho)\partial \psi_v(\rho,z)/\partial \rho$ and $H_{v\rho}(\rho,z)= -\partial \phi_v(\rho,z)/\partial \rho=(1/\rho)\partial \psi_v(\rho,z)/\partial z$, where $\nabla^2 \phi_v = 0$.  The  solutions for $z \ge 0$ are  
\begin{eqnarray}
\phi_v(\rho,z)&=&\int_0^\infty f_v(k)J_0(k\rho) e^{-kz}dk, \\
\psi_v(\rho,z)&=&-\int_0^\infty f_v(k)J_1(k\rho) \rho e^{-kz}dk,\\
H_{vz}(\rho,z)&=&\int_0^\infty f_v(k)J_0(k\rho) e^{-kz}kdk, \\
H_{v\rho}(\rho,z)&=&\int_0^\infty f_v(k)J_1(k\rho) e^{-kz}kdk, \\
f_v(k) &=& \frac{\Phi}{2\pi \mu_0}+\frac{\Phi}{\pi^2 \mu_0}g(ka,0)\\
g(ka,0)&=&-\int_a^\infty\frac{aJ_0(k\rho)}{\rho\sqrt{\rho^2-a^2}}d\rho
={\rm si}(ka)\\
{\rm si}(z)&=&-\int_z^\infty \frac{\sin t}{t}dt.
\end{eqnarray}
In the plane $z = 0$,
\begin{eqnarray}
\phi_v(\rho,0+)&=&\frac{\Phi}{\pi^2\mu_0\rho}\cos^{-1}\frac{\rho}{a},\; \rho \le a,\nonumber \\
&=&0,\; \rho \ge a,\\
\psi_v(\rho,0+)&=&-\frac{\Phi}{\pi^2\mu_0}\frac{\pi}{2},\; \rho \le a,\nonumber \\
&=&-\frac{\Phi}{\pi^2\mu_0}\sin^{-1}\frac{a}{\rho},\; \rho \ge  a,\\
H_{v\rho}(\rho,0+)&=&\frac{\Phi}{\pi^2\mu_0\rho^2}\Big(\frac{\rho}{ \sqrt{a^2-\rho^2}}+\cos^{-1}\frac{\rho}{a}\Big),\; \rho < a, \nonumber \\
&=&0,\; \rho > a,
\end{eqnarray}
and the sheet-current density in the disk ($\rho < a$) is given by $K_{v\phi}(\rho) = 2H_{v\rho}(\rho,0+)$.
The magnetic field lines generated by the flux $\Phi$ at the origin are illustrated in Fig.\ \ref{psiv}, which shows contours of the stream function $\psi_v(\rho,z)$.

\section{Small flux domes \label{Hd}}

For the case of a small flux dome of radius $b$ much smaller than the radius $a$ of the disk, it is possible to calculate the dome-generated magnetic fields $H_{sz}(\rho,z)$ and $H_{s\rho}(\rho,z)$ as well as the sheet-current density $K_{s\phi}(\rho)$ generated by vortices in the dome.  Consider the case of a disk of infinite radius, where the magnetic field generated by the vortices is
\begin{eqnarray}
H_{sz}(\rho,0) &= &H_d \sqrt{1-\rho^2/b^2}, \; \rho \le b, \label{Hdome} \\
&=&0, \; \rho \ge b.
\end{eqnarray}
Both $H_{sz}(\rho,z)$ and the corresponding radial component $H_{s\rho}(\rho,z)$ can be expressed in terms of the  magnetic potential $\phi_{s}(\rho,z)$ and stream function $\psi_s(\rho,z)$ via $H_{sz}(\rho,z)= -\partial \phi_s(\rho,z)/\partial z = -(1/\rho)\partial \psi_s(\rho,z)/\partial \rho$ and $H_{s\rho}(\rho,z)= -\partial \phi_s(\rho,z)/\partial \rho=(1/\rho)\partial \psi_s(\rho,z)/\partial z$, where $\nabla^2 \phi_s = 0$.  The  solutions for $z \ge 0$ are  
\begin{eqnarray}
\phi_s(\rho,z)&=&\int_0^\infty f_s(k)J_0(k\rho) e^{-kz}dk, \\
\psi_s(\rho,z)&=&-\int_0^\infty f_s(k)J_1(k\rho) \rho e^{-kz}dk,\\
H_{sz}(\rho,z)&=&\int_0^\infty f_s(k)J_0(k\rho) e^{-kz}kdk, \\
H_{s\rho}(\rho,z)&=&\int_0^\infty f_s(k)J_1(k\rho) e^{-kz}kdk, \\
f_s(k) &=& H_d b^2 \frac{\sin kb - kb \cos kb}{(k b)^3}.
\end{eqnarray}
In the plane $z = 0$,
\begin{eqnarray}
\phi_s(\rho,0+)&=&H_d b \frac{\pi}{4} (1-\frac{\rho^2}{2b^2}), \; \rho \le b,\\
&=& H_d b\frac{1}{4}\Big[\frac{\sqrt{\rho^2 - b^2}}{b} + (2-\frac{\rho^2}{b^2}) \sin^{-1}\frac{b}{\rho}\Big], \nonumber \\
& & \;\;\;\;\;\;\;\;\;\;\;\;\;\;\;\;\;\;\;\;\;\;\;\;\;\;\rho \ge b.\\
\psi_s(\rho,0+)
&=&-\frac{H_db^2}{3},\; \rho \le b,\nonumber \\
&=&-\frac{H_db^2}{3}\Big[1-\Big(1-\frac{\rho^2}{b^2}\Big)^{3/2}\Big],\; \rho \ge b,\\
H_{s\rho}(\rho,0+)&=& H_d \frac{\pi}{4} \frac{\rho}{b}, \; \rho \le b,\\
&=& H_d \frac{1}{2}\Big[ \frac{\rho}{b}\sin^{-1}\frac{b}{\rho}-\frac{\sqrt{\rho^2 - b^2}}{\rho} \Big], \nonumber \\
& & \;\;\;\;\;\;\;\;\;\;\;\;\;\;\rho \ge b, 
\end{eqnarray}
and the sheet-current density  is given by
\begin{equation}
K_{s\phi}(\rho)=2H_{s\rho}(\rho,0+).
\end{equation}
It can be shown that 
\begin{equation}
\int_0^\infty [K_{s\phi}(\rho)/2\rho]d\rho = H_{sz}(0,0) = H_d,
\end{equation}
as required by the Biot-Savart law.

\section{Magnetic flux $\Phi$ up through an annulus of radius $\rho_c$ \label{Hc}}

When a closely spaced array of vortices carries magnetic flux $\Phi$ up through a narrow annulus of radius $\rho_c < a$, the magnetic field and induced current density the vortices generate can be calculated as follows.   Neglecting magnetic structure on the length scale of the intervortex spacing, $\lambda$, or $\Lambda$, the $z$ component of the magnetic field generated by the vortices in the plane $z=0$ is
\begin{eqnarray}
H_{cz}(\rho,0) &= &\frac{\Phi}{2\pi\mu_0\rho}\delta(\rho-\rho_c),\; \rho < a, \label{Hcz} \\
&=&-\frac{\Phi \sqrt{a^2-\rho_c^2}}{\pi^2\mu_0(\rho^2 -\rho_c^2)\sqrt{\rho^2-a^2}}, \; \rho >a.
\end{eqnarray}
Both $H_{cz}(\rho,z)$ and the corresponding radial component $H_{c\rho}(\rho,z)$ can be expressed in terms of the  magnetic potential $\phi_{c}(\rho,z)$ and stream function $\psi_c(\rho,z)$ via $H_{cz}(\rho,z)= -\partial \phi_c(\rho,z)/\partial z = -(1/\rho)\partial \psi_c(\rho,z)/\partial \rho$ and $H_{c\rho}(\rho,z)= -\partial \phi_c(\rho,z)/\partial \rho=(1/\rho)\partial \psi_c(\rho,z)/\partial z$, where $\nabla^2 \phi_c = 0$.  The  solutions for $z \ge 0$ are  
\begin{eqnarray}
\phi_c(\rho,z)&=&\int_0^\infty f_c(k)J_0(k\rho) e^{-kz}dk, \\
\psi_c(\rho,z)&=&-\int_0^\infty f_c(k)J_1(k\rho) \rho e^{-kz}dk,\\
H_{cz}(\rho,z)&=&\int_0^\infty f_c(k)J_0(k\rho) e^{-kz}kdk,
\label{Hczrhoz} \\
H_{c\rho}(\rho,z)&=&\int_0^\infty f_c(k)J_1(k\rho) e^{-kz}kdk, 
\label{Hcrhorhoz}\\
f_c(k) &=& \frac{\Phi}{2 \pi \mu_0}J_0(k\rho_c)+\frac{\Phi}{\pi^2 \mu_0}g(ka,\rho_c/a),\\
g(ka,\rho_c/a)&=&-\int_a^\infty\frac{\rho \sqrt{a^2-\rho_c^2}J_0(k\rho)}{(\rho^2 -\rho_c^2)\sqrt{\rho^2-a^2}}d\rho.
\end{eqnarray}
In the plane $z = 0$,
\begin{eqnarray}
\phi_c(\rho,0+)&=&\frac{\Phi}{2\pi\mu_0}{\cal F}(a,\rho_c,\rho), \; \rho \le a,\label{phic} \nonumber \\
&=&0,\; \rho \ge a,\\
\psi_c(\rho,0+)&=&0,\; \rho<\rho_c,\nonumber \\
&=&-\frac{\Phi}{\pi^2\mu_0}\frac{\pi}{2},\; \rho_c<\rho \le a,\nonumber \\&=&-\frac{\Phi}{\pi^2\mu_0}\sin^{-1}\!\!\sqrt{\frac{a^2-\rho_c^2}{\rho^2-\rho_c^2}},\; \rho \ge a,\\
H_{c\rho}(\rho,0+)&=&\frac{\Phi}{2\pi\mu_0}{\cal G}(a,\rho_c,\rho),\; \rho < a, \nonumber \\
&=&0,\; \rho > a,
\end{eqnarray}
where, defining $\rho_<$ ($\rho_>$) to be the smaller (larger) of $\rho_1$ and $\rho_2$,
\begin{eqnarray}
{\cal F}(a,\rho_c,\rho)&=&F(\rho_c,\rho) \nonumber \\
&-&\frac{2}{\pi}\int_a^\infty \frac{\sqrt{a^2-\rho_c^2}F(\rho',\rho)\rho'}{(\rho'^2-\rho_c^2)\sqrt{\rho'^2-a^2}}d\rho',\\
F(\rho_1,\rho_2)&=&\int_0^\infty J_0(k\rho_1)J_0(k\rho_2)dk \\
&=&\frac{2}{\pi\rho_>}{\bm K}(\rho_</\rho_>), \\
F(\rho,0)&=&F(0,\rho)=1/\rho,\\
{\cal F}(a,\rho',0)&=&\frac{2}{\pi \rho'}\cos^{-1}\frac{\rho'}{a} ,
\label{curlyF0} \\
{\cal F}(a,\rho',a )&=&0,
\end{eqnarray}
\begin{eqnarray}
{\cal G}(a,\rho_c,\rho)&=&-\partial {\cal F}(a,\rho_c,\rho)/\partial \rho \nonumber \\
&=&G(\rho_c,\rho) \nonumber \\
&-&\frac{2}{\pi}\int_a^\infty \frac{\sqrt{a^2-\rho_c^2}G(\rho',\rho)\rho'}{(\rho'^2-\rho_c^2)\sqrt{\rho'^2-a^2}}d\rho', \\
G(\rho_1,\rho_2)&=& -\partial F(\rho_1,\rho_2)/\partial \rho_2 \\
&=&\int_0^\infty J_0(k\rho_1)J_1(k\rho_2)kdk\\
&=&\Big(\frac{1}{\pi\rho_2}\Big)\Big[\frac{{\bm K}(\kappa)}{\rho_1+\rho_2}-\frac{{\bm E}(\kappa)}{\rho_1-\rho_2}\Big]\\
&=&\Big(\frac{2}{\pi\rho_2}\Big)\frac{{\bm E}(\rho_1/\rho_2)}{\rho_2(1-\rho_1^2/\rho_2^2)},\;\rho_1<\rho_2,\\
&=&\Big(\frac{2}{\pi\rho_2}\Big)\Big[\frac{{\bm K}(\rho_2/\rho_1)}{\rho_1}-\frac{{\bm E}(\rho_2/\rho_1)}{\rho_1(1-\rho_2^2/\rho_1^2)}\Big],\nonumber \\
&&\;\;\;\;\;\;\;\;\;\;\;\;\;\;\;\;\;\;\;\;\;\;\;\;\;\;\;\;\;\;\;\;\;\;\;\rho_1>\rho_2,\\
\kappa &= &\frac{2\sqrt{\rho_1\rho_2}}{\rho_1+\rho_2},\\
G(0,\rho)&=&1/\rho^2,\\
G(\rho,0)&=&0,
\end{eqnarray} 
$J_0(x)$ and $J_1(x)$ are Bessel functions, and ${\bm K}(\kappa)$ and ${\bm E}(\kappa)$ are complete elliptic integrals of the first and second kind of modulus $\kappa$.
The sheet-current density in the disk ($\rho < a$) is given by $K_{c\phi}(\rho) = 2H_{c\rho}(\rho,0+)$.
It can be shown with the help of Eqs.\ (\ref{Hczrhoz}) and (\ref{Hcrhorhoz}) that 
\begin{equation}
\int_0^\infty[K_{c\phi}(\rho)/2\rho]d\rho = H_{cz}(0,0) = 0,
\end{equation}
as required by the Biot-Savart law.
A plot of the magnetic field lines for this case would be similar to that shown in Fig.\ \ref{psiv}, except that the field lines would emanate from $\rho = \rho_c$ rather than $\rho = 0$.

\end{document}